\documentclass[%
 aip,
 amsmath,amssymb,
 reprint,%
floatfix,
]{revtex4-1}

\usepackage[subrefformat=parens]{subcaption}
\usepackage{graphicx}
\usepackage{dcolumn}
\usepackage{bm}

\usepackage[utf8]{inputenc}
\usepackage[T1]{fontenc}
\usepackage{mathptmx}
\usepackage{multirow}
\usepackage{comment}

\begin{document}

\preprint{AIP/123-QED}

\title{Material survey for millimeter-wave absorber using 3-D printed mold}

\author{T. Otsuka}
\affiliation{
Department of Physics, Kyoto University, Kitashirakawa-Oiwakecho, Sakyo-ku, Kyoto 606-8502, Japan
}

\author{S. Adachi}
\email{shunsuke.adachi@ipmu.jp.}
\affiliation{
Kavli Institute for The Physics and Mathematics of The Universe (WPI), The University of Tokyo, Kashiwa, Chiba 277-8583, Japan
}

\author{M. Hattori}
\affiliation{
Astronomical Institute, Graduate School of Science, Tohoku University, 6-3, Aramaki Aza-Aoba, Aoba-ku, Sendai 980-8578, Japan
}

\author{Y. Sakurai}
\affiliation{
Kavli Institute for The Physics and Mathematics of The Universe (WPI), The University of Tokyo, Kashiwa, Chiba 277-8583, Japan
}

\author{O. Tajima}
\affiliation{
Department of Physics, Kyoto University, Kitashirakawa-Oiwakecho, Sakyo-ku, Kyoto 606-8502, Japan
}

\date{\today}
\begin{abstract}
Radio absorptive materials (RAMs) are key elements for receivers in the millimeter-wave range. 
For astronomical applications, cryogenic receivers are widely used to achieve a high-sensitivity. 
These cryogenic receivers, in particular the receivers for the cosmic microwave background, require that the RAM has low surface reflectance ($\lesssim 1\%$) in a wide frequency range (20--300~GHz) to minimize the undesired stray light to detectors.
We develop a RAM that satisfies this requirement based on a production technology using a 3D-printed mold (named as RAM-3pm).
This method allows us to shape periodic surface structures to achieve a low reflectance.
A wide range of choices for the absorptive materials is an advantage. 
We survey the best material for the RAM-3pm. 
We measure the index of refraction ($n$) and the extinction coefficient ($\kappa$) at liquid nitrogen temperature as well as at room temperature of 17 materials.
We also measure the reflectance at the room temperature for the selected materials.
The mixture of an epoxy adhesive (STYCAST-2850FT) and a carbon fiber (K223HE) achieves the best performance. 
We estimate the optical performance at the liquid nitrogen temperature by a simulation based on the measured $n$ and $\kappa$. 
The RAM-3pm made with this material satisfies the requirement except at the lower edge of the frequency range ($\sim$20~GHz). 
We also estimate the reflectance of a larger pyramidal structure on the surface.  %
We find a design to satisfy our requirement.
\end{abstract}

\maketitle

\section{Introduction}

Radio absorptive materials (RAMs) are indispensable elements for receivers in the millimeter-wave range. 
Their role is the mitigation of unexpected external noise (hereafter stray light), which may enter the receiver from outside of its line of sight.
Radio sensors in the receiver are generally maintained at cryogenic condition in the cases of cosmological and astronomical applications, e.g., the cosmic microwave background (CMB) observation.
In these applications, a major source of the stray light is thermal radiations from the ambient environment.
Because sensors based on a superconducting device have a high sensitivity~\cite{bib:so}, it is very important to avoid any stray light entering the sensors.

\begin{figure}[htpb]
    \centering
    \includegraphics[width=6.5cm]
    {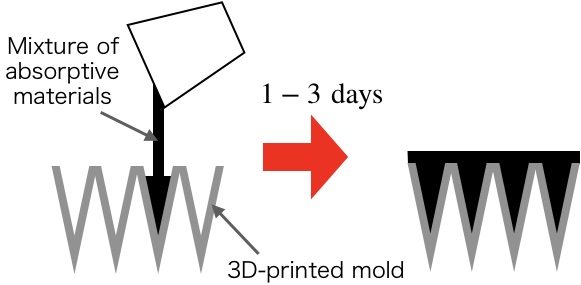}
    \caption{Production method of the RAM-3pm (radio absorptive material made with 3D-printed mold).
    The advantage of this method is the wide range of choices for the filling materials.}
    \label{HowToMake_3Dprinter_blackbody}
\end{figure}

\begin{figure}[htpb]
    \centering
    \includegraphics[width=6cm]
    {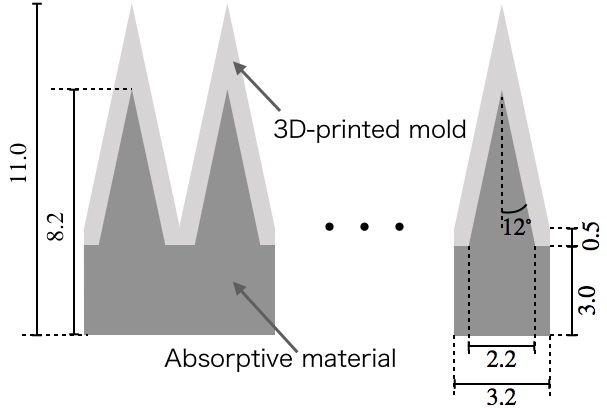}
    \\
    \vspace{0.5\baselineskip}
    \includegraphics[width=5cm]
    {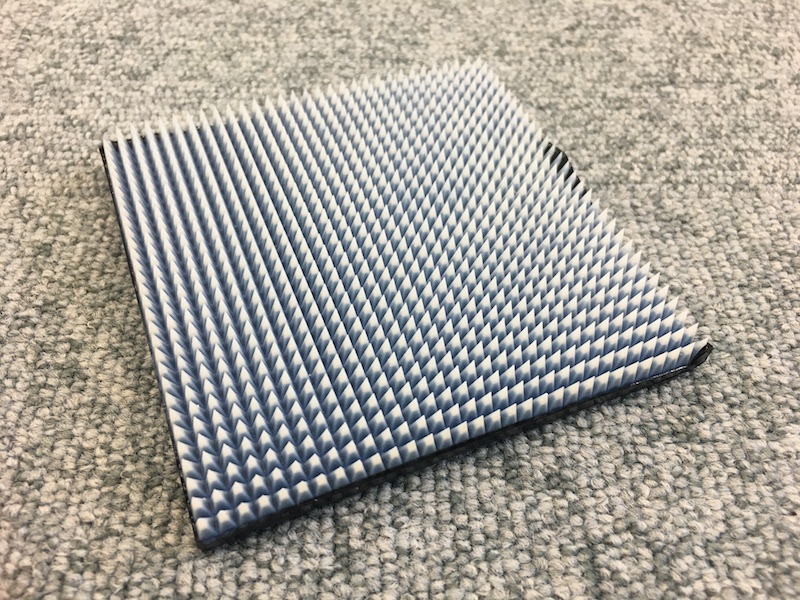}
    \caption{\label{fig:3Dprinted_RAM_photo}
    Geometry of the RAM-3pm used in this paper (top), and its photograph (bottom).
    }
\end{figure}

Shaping the periodic surface structures, e.g., a pyramid shape, is an important point in achieving a low surface reflectance. 
We previously established a production method for RAMs using a 3D-printed mold~\cite{bib:Adachi_paper}~(named as RAM-3pm).
As illustrated in Fig.~\ref{HowToMake_3Dprinter_blackbody}, we can shape the structure by filling an epoxy adhesive (STYCAST-2850FT) into a 3D-printed mold. 
Its geometry and a photograph are shown in Fig.~\ref{fig:3Dprinted_RAM_photo}.
The mold is thin (thickness of 0.5~mm) and almost transparent~\cite{bib:Adachi_paper}. 
Therefore, we do not need to remove the mold.
We also confirmed that the mold is usable for cryogenic application. 
A wide range of choices for the filling material is an important advantage of the RAM-3pm compared with the other RAMs~\cite{bib:MMA,bib:EdWollack}. 

In this paper, we survey the best material for the RAM-3pm. 
We focus on the CMB application, which requires a low surface reflectance ($\lesssim 1\%$) in a wide frequency range~(20--300~GHz) in a cryogenic condition~\cite{bib:so}.
In section~II, we prepare 17 materials. 
In section~III, we make flat-shaped samples, which are a mixture of each material and STYCAST-2850FT, 
and we measure their optical properties: index of refraction and extinction coefficient.
To evaluate the performance at low temperature, we perform these measurements at liquid nitrogen temperature (77~K) as well as at room temperature ($\sim 290$~K).
In section~IV, we select candidate materials based on these optical parameters, and we make a RAM-3pm for each of them.
Their surface reflectance is measured at room temperature.
We check the consistency between the measurement and simulation using the measured optical parameters.
In section~V, we estimate their performance at 77~K including their scattering effects.
We provide our conclusions in section~VI.

\section{Materials}

\begin{figure}[htpb]
    \centering
    \includegraphics[width=\linewidth]
    {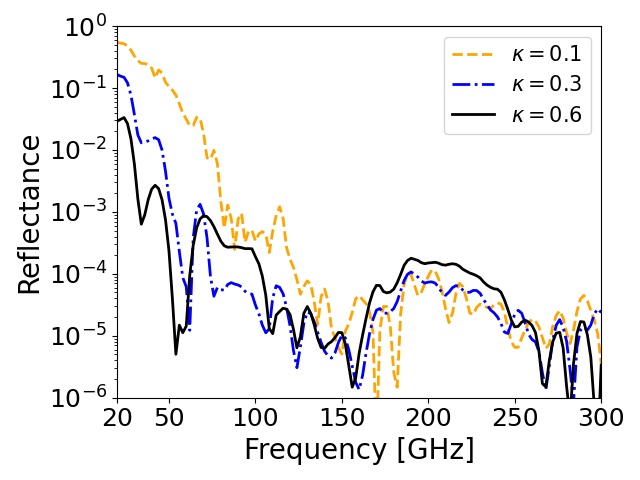}
    \caption{Simulated reflectance for the RAM whose geometry is defined in Fig.~\ref{fig:3Dprinted_RAM_photo}. 
    We simulate three cases: $\kappa$~=~0.1, 0.3, and 0.6 with a fixed $n=2.3$.
    In the low frequency range, we obtain a lower reflectance with a larger $\kappa$.
    }
    \label{k_comparison}
\end{figure}

The surface reflectance of the RAM relies on the surface structure and optical properties of the material, i.e., the index of refraction~($n$) and extinction coefficient~($\kappa$). 
Figure~\ref{k_comparison} shows the simulation results for three cases: 
$(n, \kappa)$~=~(2.3, 0.1), (2.3, 0.3), and (2.3, 0.6)
for the RAM whose geometry is defined in Fig.~\ref{fig:3Dprinted_RAM_photo}.
We used ANSYS-HFSS\cite{bib:ANSYS} for the simulation, and we assumed unpolarized light whose incident angle was 45$^{\circ}$.
We set a perfect conductor at the backside of the RAM in the simulation.
We found that a lower reflectance is obtained with a larger $\kappa$ in the low frequency range, where the reflectance is relatively high.
We can intuitively understand this by the formula that describes light attenuation: Lambert's law~\cite{bib:jackson},
\begin{equation}
    I = I_0 \exp \left(- \frac{4 \pi \nu \kappa d}{c} \right), \label{eq:Lambert}
\end{equation}
where 
$I$ is the output power through the material,
$I_0$ is the power of input light, 
$\nu$ is the frequency,
$d$ is the thickness of the optical path, 
and $c$ is the speed of light. 
Finding a material that has a large $\kappa$ is our strategy to achieve our requirement.

\begin{figure*}[htpb]
    \centering
    \begin{minipage}{0.24\textwidth}
        \begin{center}
            \includegraphics[width=0.9\textwidth]
            {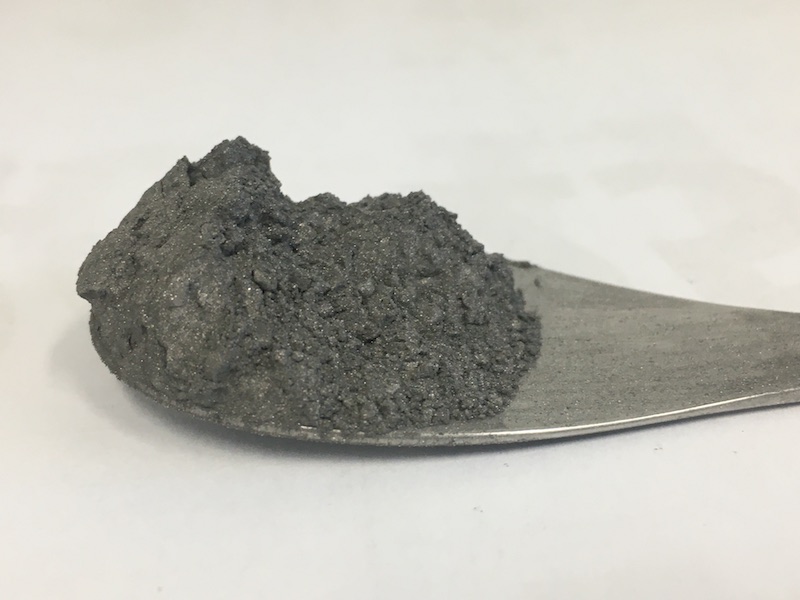}
            \subcaption{Stainless steel (RD19-3532)}
        \end{center}
    \end{minipage}
    \begin{minipage}{0.24\textwidth}
        \begin{center}
            \includegraphics[width=0.9\textwidth]
            {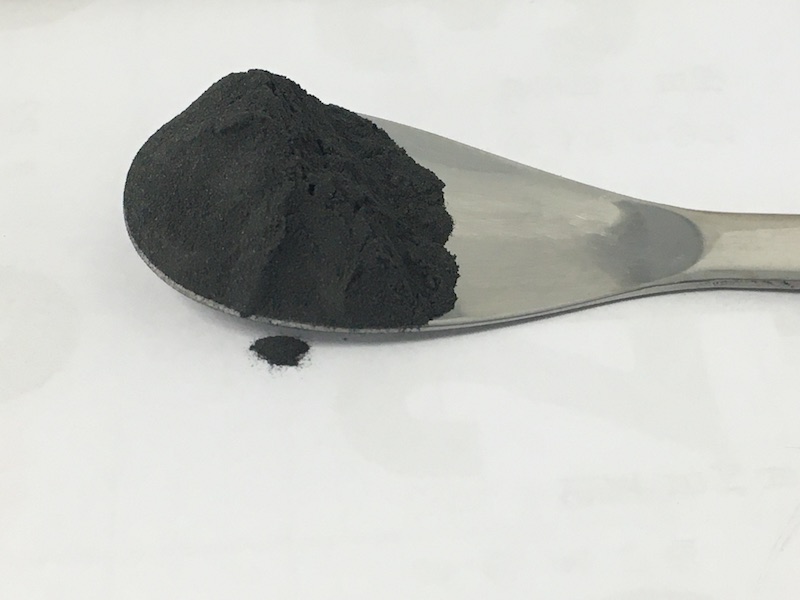}
            \subcaption{Graphite (SG-BL40)}
        \end{center}
    \end{minipage}
    \begin{minipage}{0.24\textwidth}
        \begin{center}
            \includegraphics[width=0.9\textwidth]
            {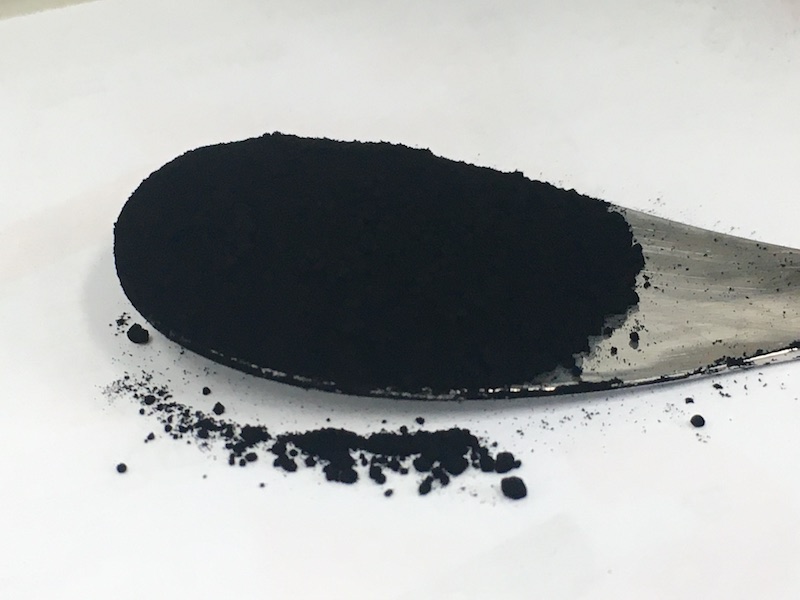}
            \subcaption{Carbon black (\#3230B)}
        \end{center}
    \end{minipage}
    \begin{minipage}{0.24\textwidth}
        \begin{center}
            \includegraphics[width=0.9\textwidth]
            {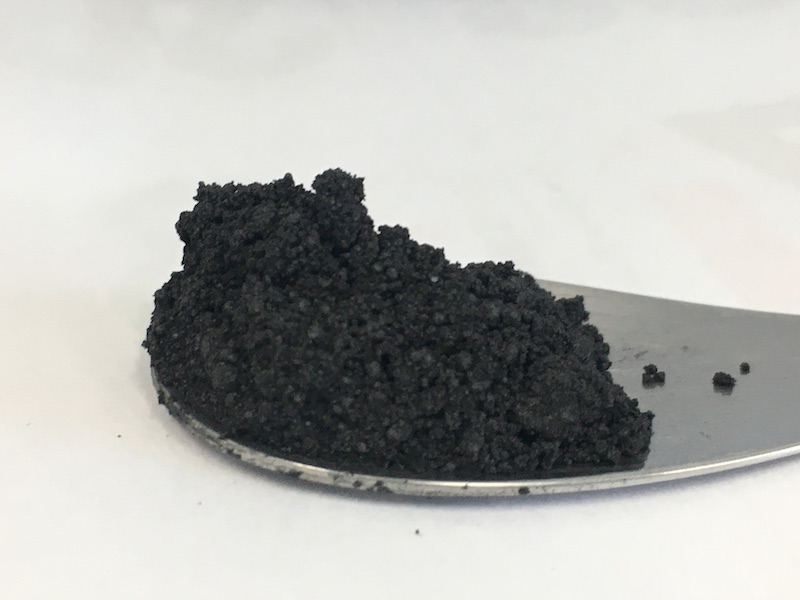}
            \subcaption{Carbon fiber (CNF)}
        \end{center}
    \end{minipage} 
    \\
    \begin{minipage}{0.24\textwidth}
        \begin{center}
            \includegraphics[width=0.9\textwidth]
            {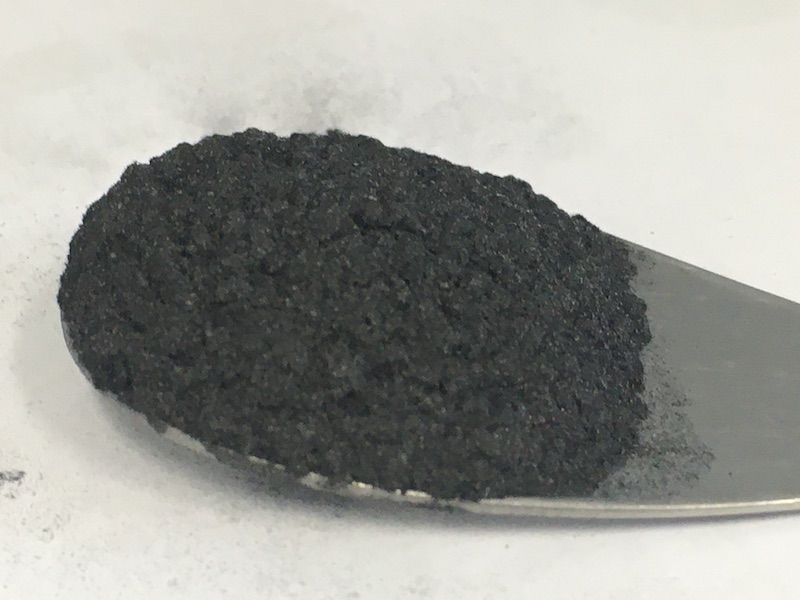}
            \subcaption{Carbon fiber (K223HM)}
        \end{center}
    \end{minipage}
    \begin{minipage}{0.24\textwidth}
        \begin{center}
            \includegraphics[width=0.9\textwidth]
            {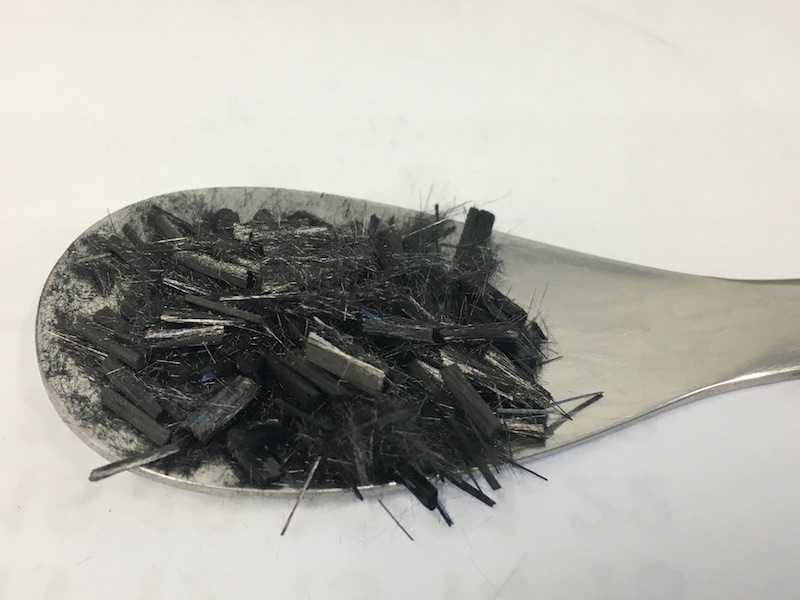}
            \subcaption{Carbon fiber (K223HE)}
        \end{center}
    \end{minipage}
    \begin{minipage}{0.24\textwidth}
        \begin{center}
            \includegraphics[width=0.9\textwidth]
            {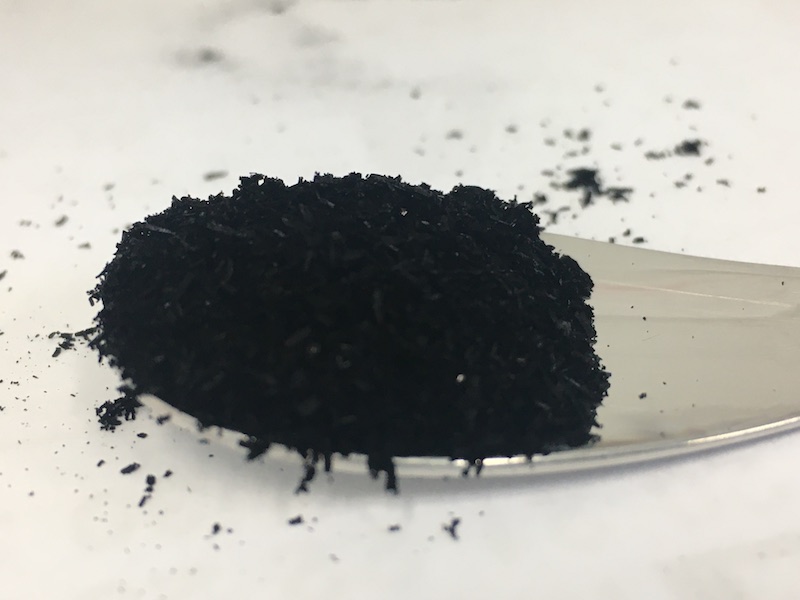}
            \subcaption{Carbon nanotube (SG101)}
        \end{center}
    \end{minipage}
       \begin{minipage}{0.24\textwidth}
        \begin{center}
            \includegraphics[width=0.7\textwidth]
            {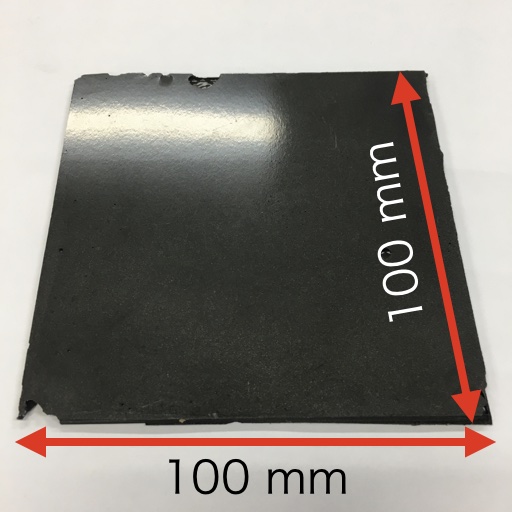}
            \subcaption{Flat-shaped sample}
        \end{center}
    \end{minipage}
    \caption{(a)-(g) Photos of absorptive materials listed in Table~\ref{comparison_powders}. 
    (h) One of the flat-shaped samples, a mixture of stainless steel powder RD19-3532 and STYCAST-2850FT.}
    \label{powder_photo}
\end{figure*}

We used mixtures of a base material and absorptive materials.
We selected STYCAST-2850FT\cite{bib:Henkel} as the base material
because it has been widely used for cryogenic applications. 
The RAM-3pm made with this material has a sufficient mechanical strength and a good thermal conductivity in the cryogenic condition according to our previous study~\cite{bib:Adachi_paper}. 
For the materials mixed into the STYCAST-2850FT, we prepared stainless steel powders and carbon materials because they are expected to have a large $\kappa$ owing to their electrical conductivity. 
These are listed in Table~\ref{comparison_powders}.
This table includes their product names, their particle sizes, and the weight fractions of each mixture sample.
We surveyed 17 materials in total.
Photos of seven of them are shown in Fig.~\ref{powder_photo}. 
The weight fraction was determined to maintain a low viscosity for  easy fabrication of the RAM-3pm.
It was approximately $6~\mathrm{Pa}\cdot\mathrm{s}$ in the case of the carbon fiber K223HE.

\begin{table*}[htpb]
    \caption{Absorptive materials mixed with STYCAST-2850FT. 
    The carbon fibers and carbon nanotube were cylindrical particles while the others were spherical or flat particles.
    The size of the cylindrical particles was defined by the length ($L$) and the diameter ($\phi$). 
    We made flat-shaped samples for measurements of $n$ and $\kappa$. 
    Each sample was mixed into STYCAST-2850FT with the listed weight fraction.
    }\label{comparison_powders}
    \begin{ruledtabular}
    \begin{tabular}{@{~~~~}lllcc r@{~~~~}} 
        \multicolumn{2}{c}{Types of materials} & \multicolumn{1}{c}{Product name}   & \multicolumn{1}{c}{Particle size} & Photo index at Fig.~\ref{powder_photo} & \multicolumn{1}{c}{Weight fraction} \\ 

        \hline
        Stainless steel\cite{bib:ToyoAluminium} & Spherical shape & RD19-3532 & {$10 ~ \mu \rm m$} & (a) & 20\% \\
        & Flat shape & RFA6500 &  {$17 ~ \mu \rm m$} & -- & 7\% \\ 

        \hline
        Graphite \cite{bib:ItoGraphite}
        & Spherical graphite  & SG-BL40 & {$38 ~ \mu \rm m$} & (b) & 14\% \\ 
        & Flake graphite & Z-50 & {$48 ~ \mu \rm m$} & -- &4\% \\ 
        & Pyrolytic graphite & PC99-300M & {$38 ~ \mu \rm m$} & -- & 3.5\% \\ 
        & Expanded graphite & EC300 & {$39 ~ \mu \rm m$} & -- & 1\% \\ 
        & Artificial graphite & AGB-604  & {$44 ~ \mu \rm m$} & -- & 5\% \\ 

        \hline
        Carbon black \cite{bib:MCCarbonBlack}
        & Conductive furnace & \#3230B & {$23 ~ \rm nm$} & (c) & 0.7\% \\ 
        & Long flow furnace & MA230 & {$30 ~ \rm nm$} & -- & 0.7\% \\ 
        & \multirow{2}{*}{Regular color furnace} & \#10 & {$75 ~ \rm nm$} & -- & 2\% \\ 
        & & \#95 & {$40 ~ \rm nm$} & -- & 1.2\% \\ 
        & Medium color furnace & MA600 & {$20 ~ \rm nm$} & -- & 0.7\% \\ 
        & High color furnace & \#2600 & {$13 ~ \rm nm$} & -- & 0.5\% \\ 

        \hline
        Carbon fiber 
        & Nano fiber\cite{bib:ALMEDIO} & CNF & $L 1 \sim 20 ~ \mu \rm m$,~~$\phi 200 \sim 800 ~ \rm nm$ & (d) & 0.7\% \\ 
        & Milled fiber\cite{bib:MCCarbonFiber} & K223HM & $L 50 \sim 200 ~ \mu \rm m$, ~~ $\phi 11 ~ \mu \rm m$ & (e) & 2\% \\
        & Chopped fiber\cite{bib:MCCarbonFiber} & K223HE & $L 6 ~ \rm mm$, ~~ $\phi 11 ~ \mu \rm m$ & (f) & 0.15\% \\ 

        \hline
        \multicolumn{2}{l}{Carbon nanotube\cite{bib:ZeonNanoTechnology}} 
        & SG101 & $L 100 \sim 600 ~ \mu \rm m$, ~~ $\phi 3 \sim 5 ~ \rm nm$ & (g) & 0.2\% \\ 
    \end{tabular}
    \end{ruledtabular}
\end{table*}

\section{Optical parameters}

\begin{figure*}[htpb]
    \centering
    \begin{minipage}{0.33\textwidth}
        \begin{center}
            \includegraphics[width=0.95\textwidth]
            {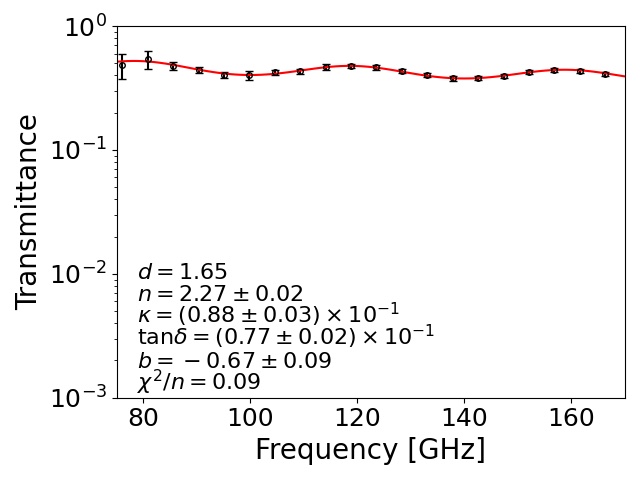}
            \subcaption{STYCAST-2850FT (no mixture)}
        \end{center}
    \end{minipage}
    \begin{minipage}{0.33\textwidth}
        \begin{center}
            \includegraphics[width=0.95\textwidth]
            {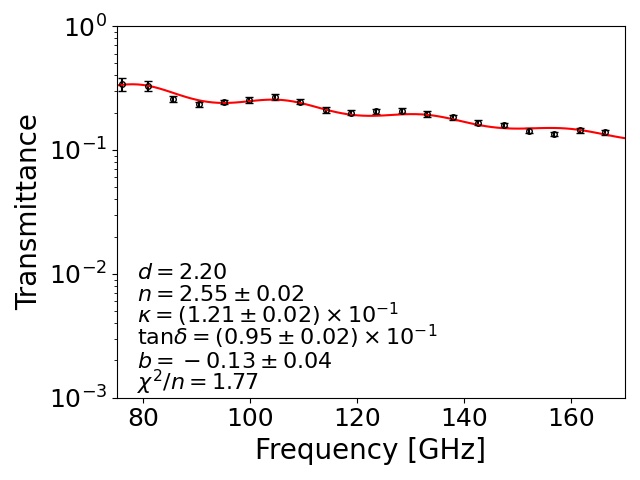}
            \subcaption{Stainless steel (RD19-3532, 20\%)}
        \end{center}
    \end{minipage}
    \begin{minipage}{0.33\textwidth}
        \begin{center}
            \includegraphics[width=0.95\textwidth]
            {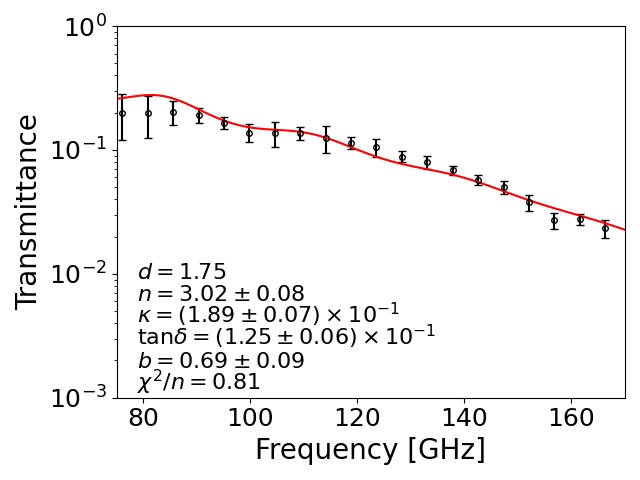}
            \subcaption{Graphite (Z-50, 4\%)}
        \end{center}
    \end{minipage} \\
    \begin{minipage}{0.33\textwidth}
        \begin{center}
            \includegraphics[width=0.95\textwidth]
            {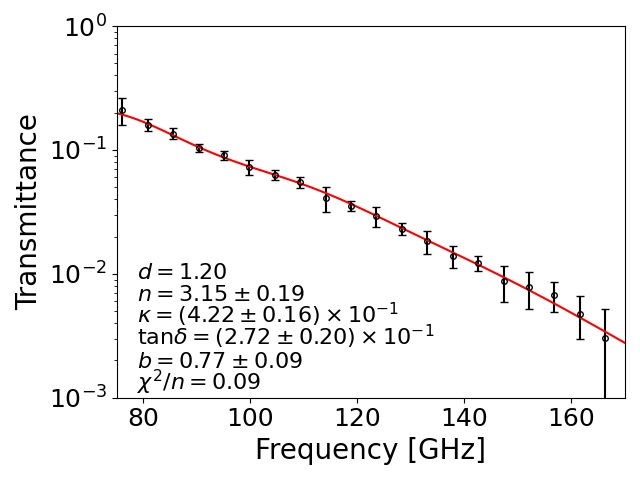}
            \subcaption{Carbon fiber (K223HM, 2\%)}
        \end{center}
    \end{minipage}
    \begin{minipage}{0.33\textwidth}
        \begin{center}
            \includegraphics[width=0.95\textwidth]
            {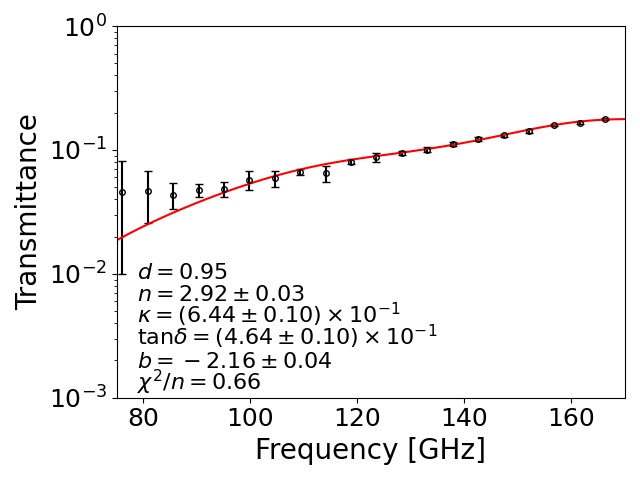}
            \subcaption{Carbon fiber (K223HE, 0.15\%)}
        \end{center}
    \end{minipage}
    \begin{minipage}{0.33\textwidth}
        \begin{center}
            \includegraphics[width=0.95\textwidth]
            {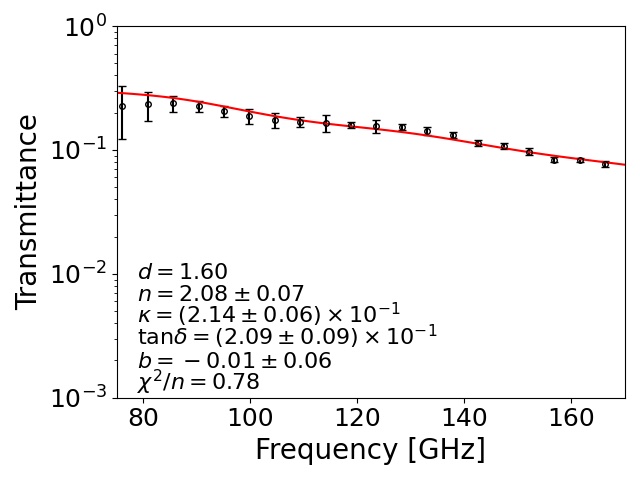}
            \subcaption{Carbon nanotube (SG101, 0.2\%)}
        \end{center}
    \end{minipage}
    \caption{Transmittance as a function of frequency at liquid nitrogen temperature. The percentage in each sub-caption is the mixing weight fraction for each sample.}
    \label{Transmittance_somesamples}
\end{figure*}

\begin{figure*}[htpb]
    \includegraphics[width=0.8\textwidth]
    {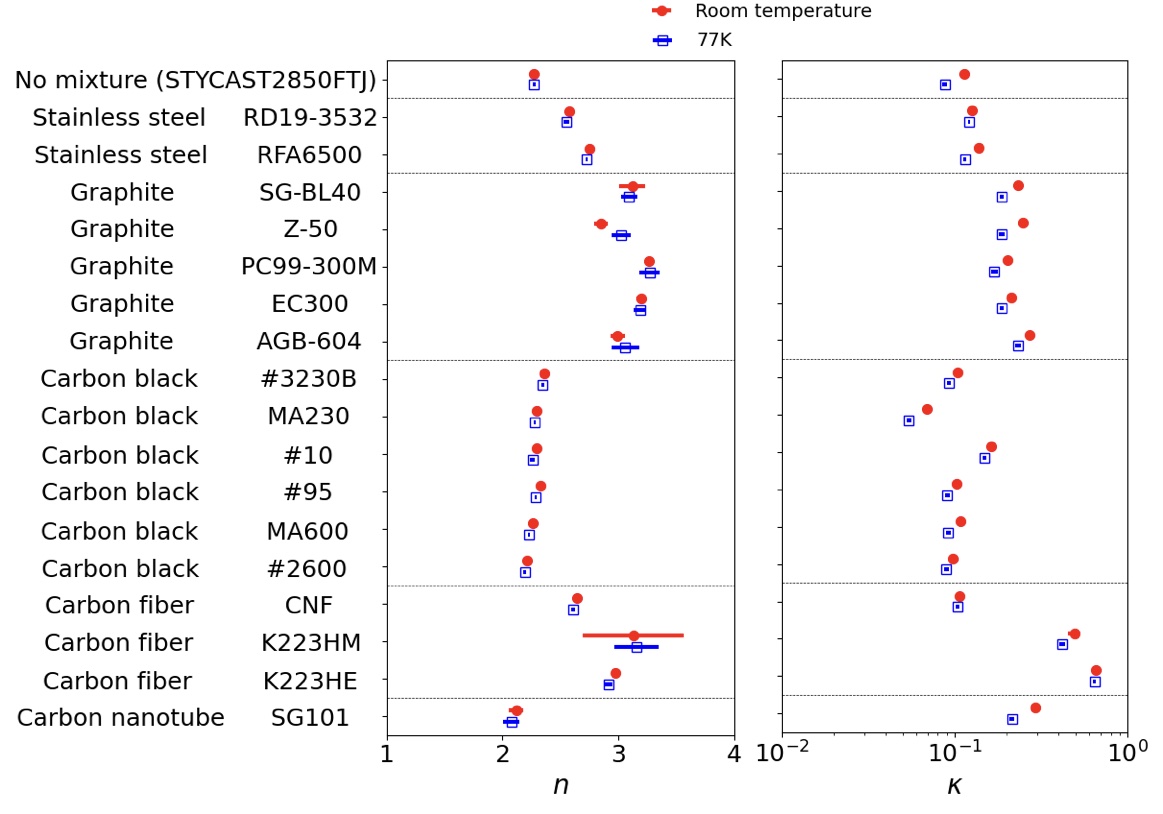}
    \caption{Optical parameters ($n$ and $\kappa$) obtained from the measured transmittance of the flat-shaped samples at 77 K and room temperature.}
    \label{n_kappa_allsamples}
\end{figure*}

\begin{table*}[!htpb]
\caption{\label{table:nk_stycast}
    Optical parameters obtained from the transmittance of the flat-shaped samples:
    index of refraction~($n$), extinction coefficient~($\kappa$), and frequency dependence~($b$).
    Their definitions are described in Eqs.~(\ref{eq:T}).
    Each material is mixed into STYCAST-2850FT with the listed weight fraction.
    We also show the result of the no-mixture sample as a reference.
    Note that the dielectric loss-tangent is calculated from the $n$ and $\kappa$, 
    e.g., $\tan\delta = 2n \kappa /(n^2 - \kappa^2) = 0.46$ for the K223HE at 77~K.
    }
    \begin{ruledtabular}
    \begin{tabular}{@{~~~~}lclrr@{~~~~}} 
        {Mixed material (product name)} &
        Weight fraction & 
        \multicolumn{1}{c}{Optical parameter} & \multicolumn{1}{c}{Room temperature} & \multicolumn{1}{c}{77~K} \\
    \hline
    \multirow{3}{*}{No mixture (STYCAST-2850FT)}
        & \multirow{3}{*}{--} & $n$ & $~~2.27 \pm 0.01 \pm 0.01$
        & $~~2.27 \pm 0.02 \pm 0.01$ \\
        & & $\kappa ~ [\times 10^{-1}]$ & $~~1.13 \pm 0.02 \pm 0.82$
        & $~~0.88 \pm 0.03 \pm 0.80$ \\
        & & $b$ & $-0.56 \pm 0.04 \pm 1.05$
        & $-0.67 \pm 0.09 \pm 2.10$ \\
        
    \hline
    \multirow{3}{*}{Stainless steel (RD19-3532)}
        & \multirow{3}{*}{20\%} & $n$ & $~~2.57 \pm 0.01 \pm 0.01$
        & $~~2.55 \pm 0.02 \pm 0.03$ \\
        & & $\kappa ~ [\times 10^{-1}]$ & $~~1.26 \pm 0.01 \pm 0.58$
        & $~~1.21 \pm 0.02 \pm 0.59$ \\
        & & $b$ & $-0.17 \pm 0.03 \pm 0.54$
        & $-0.13 \pm 0.04 \pm 0.57$ \\
        
    \hline
    \multirow{3}{*}{Stainless steel (RFA6500)}
        & \multirow{3}{*}{7\%} & $n$ & $~~2.75 \pm 0.01 \pm 0.01$
        & $~~2.72 \pm 0.01 \pm 0.01$ \\
        & & $\kappa ~ [\times 10^{-1}]$ & $~~1.37 \pm 0.02 \pm 0.99$
        & $~~1.15 \pm 0.02 \pm 1.00$ \\
        & & $b$ & $-0.35 \pm 0.03 \pm 1.25$
        & $-0.41 \pm 0.06 \pm 2.06$ \\
        
    \hline
    \multirow{3}{*}{Graphite (SG-BL40)}
        & \multirow{3}{*}{14\%} & $n$ & $~~3.12 \pm 0.12 \pm 0.01$
        & $~~3.09 \pm 0.07 \pm 0.01$ \\
        & & $\kappa ~ [\times 10^{-1}]$ & $~~2.33 \pm 0.05 \pm 0.54$
        & $~~1.88 \pm 0.06 \pm 0.53$ \\
        & & $b$ & $-0.12 \pm 0.06 \pm 0.21$
        & $-0.06 \pm 0.07 \pm 0.29$ \\

    \hline
    \multirow{3}{*}{Graphite (Z-50)}
        & \multirow{3}{*}{4\%} & $n$ & $~~2.85 \pm 0.06 \pm 0.01$
        & $~~3.02 \pm 0.08 \pm 0.02$ \\
        & & $\kappa ~ [\times 10^{-1}]$ & $~~2.50 \pm 0.05 \pm 0.75$
        & $~~1.89 \pm 0.07 \pm 0.74$ \\
        & & $b$ & $~~0.20 \pm 0.05 \pm 0.37$
        & $~~0.69 \pm 0.09 \pm 0.66$ \\

    \hline
    \multirow{3}{*}{Graphite (PC99-300M)}
        & \multirow{3}{*}{3.5\%} & $n$ & $~~3.27 \pm 0.02 \pm 0.01$
        & $~~3.27 \pm 0.09 \pm 0.83$ \\
        & & $\kappa ~ [\times 10^{-1}]$ & $~~2.03 \pm 0.03 \pm 0.75$
        & $~~1.70 \pm 0.07 \pm 0.43$ \\
        & & $b$ & $~~0.25 \pm 0.04 \pm 0.52$
        & $~~0.63 \pm 0.10 \pm 0.36$ \\

    \hline
    \multirow{3}{*}{Graphite (EC300)}
        & \multirow{3}{*}{1\%} & $n$ & $~~3.20 \pm 0.04 \pm 0.02$
        & $~~3.19 \pm 0.05 \pm 0.86$ \\
        & & $\kappa ~ [\times 10^{-1}]$ & $~~2.14 \pm 0.04 \pm 0.73$
        & $~~1.88 \pm 0.04 \pm 0.40$ \\
        & & $b$ & $~~0.80 \pm 0.06 \pm 0.63$
        & $~~0.99 \pm 0.06 \pm 0.36$ \\

    \hline
    \multirow{3}{*}{Graphite (AGB-604)}
        & \multirow{3}{*}{5\%} & $n$ & $~~2.99 \pm 0.06 \pm 0.01$
        & $~~3.06 \pm 0.12 \pm 0.01$ \\
        & & $\kappa ~ [\times 10^{-1}]$ & $~~2.73 \pm 0.05 \pm 0.79$
        & $~~2.34 \pm 0.10 \pm 0.79$ \\
        & & $b$ & $~~0.25 \pm 0.05 \pm 0.35$
        & $~~0.47 \pm 0.11 \pm 0.50$ \\

    \hline
    \multirow{3}{*}{Carbon black (\#3230B)}
        & \multirow{3}{*}{0.7\%} & $n$ & $~~2.36 \pm 0.01 \pm 0.01$
        & $~~2.35 \pm 0.01 \pm 0.01$ \\
        & & $\kappa ~ [\times 10^{-1}]$ & $~~1.05 \pm 0.02 \pm 0.68$
        & $~~0.93 \pm 0.03 \pm 0.72$ \\
        & & $b$ & $-0.71 \pm 0.04 \pm 0.07$
        & $-0.73 \pm 0.08 \pm 0.28$ \\

    \hline
    \multirow{3}{*}{Carbon black (MA230)}
        & \multirow{3}{*}{0.7\%} & $n$ & $~~2.29 \pm 0.01 \pm 0.01$
        & $~~2.28 \pm 0.01 \pm 0.01$ \\
        & & $\kappa ~ [\times 10^{-1}]$ & $~~0.69 \pm 0.01 \pm 0.50$
        & $~~0.55 \pm 0.02 \pm 0.51$ \\
        & & $b$ & $-0.44 \pm 0.04 \pm 0.85$
        & $-0.58 \pm 0.09 \pm 2.72$ \\

    \hline
    \multirow{3}{*}{Carbon black (\#10)}
        & \multirow{3}{*}{2\%} & $n$ & $~~2.29 \pm 0.01 \pm 0.01$
        & $~~2.26 \pm 0.02 \pm 0.01$ \\
        & & $\kappa ~ [\times 10^{-1}]$ & $~~1.63 \pm 0.02 \pm 0.73$
        & $~~1.49 \pm 0.04 \pm 0.76$ \\
        & & $b$ & $-0.70 \pm 0.03 \pm 0.15$
        & $-0.76 \pm 0.06 \pm 0.24$ \\

    \hline
    \multirow{3}{*}{Carbon black (\#95)}
        & \multirow{3}{*}{1.2\%} & $n$ & $~~2.33 \pm 0.01 \pm 0.01$
        & $~~2.29 \pm 0.01 \pm 0.01$ \\
        & & $\kappa ~ [\times 10^{-1}]$ & $~~1.02 \pm 0.02 \pm 0.74$
        & $~~0.91 \pm 0.03 \pm 0.80$ \\
        & & $b$ & $-0.77 \pm 0.04 \pm 0.41$
        & $-0.98 \pm 0.08 \pm 0.80$ \\

    \hline
    \multirow{3}{*}{Carbon black (MA600)}
        & \multirow{3}{*}{0.7\%} & $n$ & $~~2.27 \pm 0.01 \pm 0.01$
        & $~~2.23 \pm 0.01 \pm 0.01$ \\
        & & $\kappa ~ [\times 10^{-1}]$ & $~~1.08 \pm 0.01 \pm 0.73$
        & $~~0.92 \pm 0.03 \pm 0.76$ \\
        & & $b$ & $-0.48 \pm 0.04 \pm 0.73$
        & $-0.65 \pm 0.08 \pm 1.33$ \\

    \hline
    \multirow{3}{*}{Carbon black (\#2600)}
        & \multirow{3}{*}{0.5\%} & $n$ & $~~2.21 \pm 0.01 \pm 0.01$
        & $~~2.19 \pm 0.01 \pm 0.01$ \\
        & & $\kappa ~ [\times 10^{-1}]$ & $~~0.98 \pm 0.01 \pm 0.73$
        & $~~0.90 \pm 0.03 \pm 0.79$ \\
        & & $b$ & $-0.59 \pm 0.04 \pm 0.87$
        & $-0.96 \pm 0.10 \pm 0.94$ \\

    \hline
    \multirow{3}{*}{Carbon nanofiber (CNF)}
        & \multirow{3}{*}{0.7\%} & $n$ & $~~2.64 \pm 0.01 \pm 0.01$
        & $~~2.61 \pm 0.02 \pm 0.01$ \\
        & & $\kappa ~ [\times 10^{-1}]$ & $~~1.07 \pm 0.01 \pm 0.63$
        & $~~1.04 \pm 0.03 \pm 0.64$ \\
        & & $b$ & $-0.05 \pm 0.03 \pm 0.83$
        & $~~0.24 \pm 0.07 \pm 1.09$ \\

    \hline
    \multirow{3}{*}{Carbon fiber (K223HM)}
        & \multirow{3}{*}{2\%} & $n$ & $~~3.13 \pm 0.44 \pm 0.07$
        & $~~3.15 \pm 0.19 \pm 0.12$ \\
        & & $\kappa ~ [\times 10^{-1}]$ & $~~4.95 \pm 0.42 \pm 1.07$
        & $~~4.22 \pm 0.16 \pm 1.04$ \\
        & & $b$ & $~~0.60 \pm 0.19 \pm 0.33$
        & $~~0.77 \pm 0.09 \pm 0.41$ \\

    \hline
    \multirow{3}{*}{Carbon fiber (K223HE)}
        & \multirow{3}{*}{0.15\%} & $n$ & $~~2.97 \pm 0.02 \pm 1.05$
        & $~~2.92 \pm 0.03 \pm 0.01$ \\
        & & $\kappa ~ [\times 10^{-1}]$ & $~~6.57 \pm 0.11 \pm 0.32$
        & $~~6.44 \pm 0.10 \pm 1.44$ \\
        & & $b$ & $-2.11 \pm 0.04 \pm 0.22$
        & $-2.16 \pm 0.04 \pm 0.39$ \\

    \hline
    \multirow{3}{*}{Carbon nanotube (SG101)}
        & \multirow{3}{*}{0.2\%} & $n$ & $~~2.12 \pm 0.06 \pm 0.01$
        & $~~2.08 \pm 0.07 \pm 0.02$ \\
        & & $\kappa ~ [\times 10^{-1}]$ & $~~2.92 \pm 0.05 \pm 0.83$
        & $~~2.14 \pm 0.06 \pm 0.83$ \\
        & & $b$ & $-0.23 \pm 0.03 \pm 0.25$
        & $-0.01 \pm 0.06 \pm 0.43$ \\
    \end{tabular}
    \end{ruledtabular}
\end{table*}

To understand $n$ and $\kappa$ for each sample, we made flat-shaped samples, as shown in Fig.~\ref{powder_photo}~(h). 
We measured their transmittance at low temperature as well as at room temperature. 
We used a Martin--Puplett-type Fourier transform spectrometer (FTS) with a semiconducting bolometer~\cite{bib:HattoriFTS,bib:HattoriBolometer}.
The transmittance was obtained by the ratio of measured powers with and without the sample.
For the low-temperature measurements, we took the ratio of the data for a liquid nitrogen bath itself without the sample and the data for the sample set in this bath.
Figure~\ref{Transmittance_somesamples} shows the transmittance at 77~K for STYCAST-2850FT, and the samples doped with additional absorptive materials:~\
stainless steel, graphite, two types of carbon fibers, and carbon nanotube.

The transmittance is modeled with the following formulas,
\begin{subequations} \label{eq:T}
\begin{eqnarray}
    T (\nu) &=& \left|\frac{\tau \exp\left(-i  \frac{2 \pi \nu d }{c}N \right)}
    {1 - \rho \exp\left(-i  \frac{4 \pi \nu d}{c} N \right)}\right|^2
    , 
    \\
    \tau &=& 
    \frac{2 N}{N + n_0}
    \times
    \frac{2 n_0}{N + n_0}
    ,
    \\
    \rho &=& \left(
            \frac{ N - n_0}{N + n_0}
            \right)^2
    ,            
    \\
    N &=& n - i \kappa \left(\frac{\nu}{\nu_{100}}\right)^b
    ~(\nu_{100} = 100~{\rm GHz})
    .
\end{eqnarray}
\end{subequations}
Here, $N$ is the complex index of refraction for each sample.
The $n_0$ is an ordinary index of refraction of air or liquid nitrogen.
Note that the extinction coefficients of air and liquid nitrogen are negligible.
We allowed for the fact that the $\kappa$ has a frequency dependence by adding the parameter $b$.
For $n$, we did not observe any frequency dependence within the accuracy of our measurements.
We extracted $n$, $\kappa$, and $b$ by a fitting to the measured transmittance using the above formulas.
We fixed the thickness ($d$) because it was measured by using a caliper for each sample with an accuracy of $0.05~\mathrm{mm}$.
We also fixed $n_0 = 1.00~(1.20)$ for the room temperature (liquid nitrogen temperature) measurement.

Extracted optical parameters for each sample at each temperature condition are summarized in Fig.~\ref{n_kappa_allsamples} and Table~\ref{table:nk_stycast}.
We found that the two samples that contained the carbon fibers, K223HE and K223HM, achieved a large $\kappa$.
These were significantly higher than the sample using stainless steel (RD19-3532), which is currently used for the CMB project~\cite{bib:so}.
There was no significant difference for $n$ between the two temperature conditions.
However, the $\kappa$ at 77~K was lower than the value at room temperature for most of the samples.
The difference of $\kappa$ between 77~K and room temperature was small ($< 4\%$) for three samples:~\
the stainless steel powder (RD19-3532) and two types of carbon fibers (CNF and K223HE).

\begin{figure}[htpb]
    \centering        
        \includegraphics[width=7cm]
        {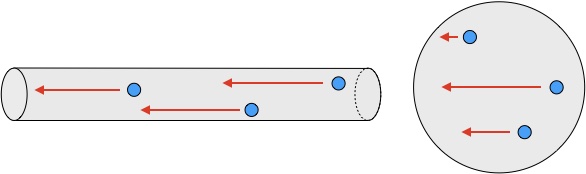}
        \caption{The mean free path of the electron relies on the particle shape. The long cylindrical particle has a longer mean free path than the spherical particle.
        }
        \label{flow2}
\end{figure}

In Table~\ref{table:nk_stycast}, extracted parameters had two types of error: statistical uncertainty (middle values for each row) and systematic uncertainty (right values for each row).
The statistical errors were assigned with the standard deviations of 20~times the FTS measurements.
For several samples, we also measured the transmittance using a vector network analyzer~(VNA). 
We observed a difference of linearity for the response between the FTS measurement and the VNA measurement.
Therefore, we conservatively assigned the effect of this linearity difference as the systematic error.  
The sample whose particle size was large tended to have a high $\kappa$.
One of possible interpretations is that $\kappa$ is related to the mean free path of the free electrons in the material.
The absorption of radio waves is dependent on the electrical conductivity of the material. 
The sample doped with large particles had a long mean free path of the electron.
In particular, the two carbon fibers, K223HM and K223HE, had a large $\kappa$.
This is because the particle shape of these fibers is a long cylinder, and the mean free path of the electron is significantly longer than the cases with the other materials (Fig.~\ref{flow2}).
The carbon nanotubes are predisposed to clumping. 
This property resulted in a shorter mean free path of the electron than for the cases of the carbon fibers. 
 
\section{Reflectance of selected materials}

\begin{figure*}[htpb]
    \centering
    \begin{minipage}{0.45\textwidth}
            \includegraphics[width=0.9\textwidth]
            {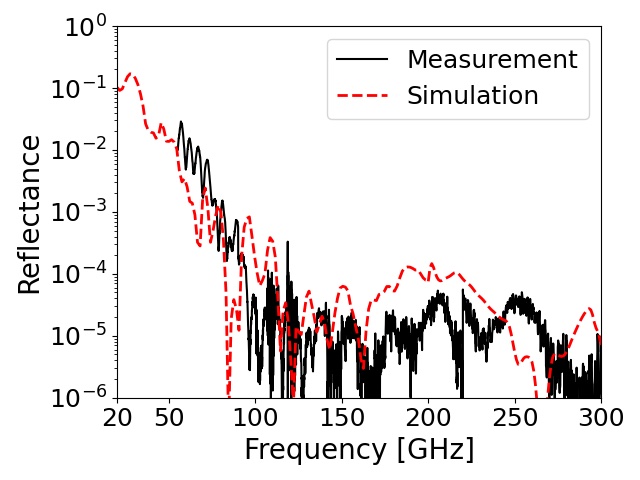}
            \subcaption{Stainless steel(RD19-3532, 20\%)}
    \end{minipage}
    \begin{minipage}{0.45\textwidth}
            \includegraphics[width=0.9\textwidth]
            {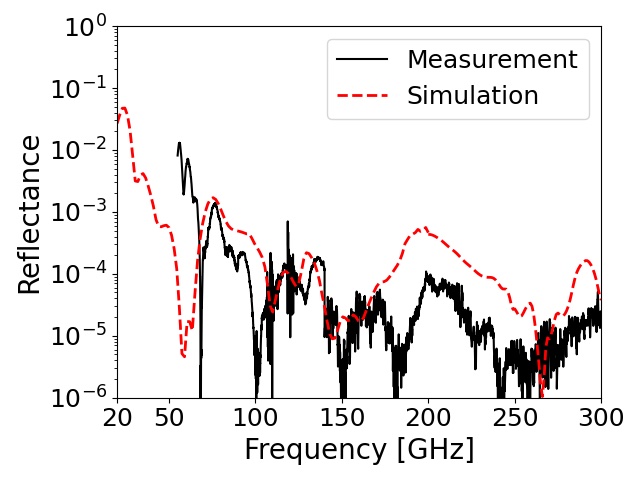}
            \subcaption{Carbon fiber (K223HM, 2\%)}
    \end{minipage}
    \begin{minipage}{0.45\textwidth}
            \includegraphics[width=0.9\textwidth]
            {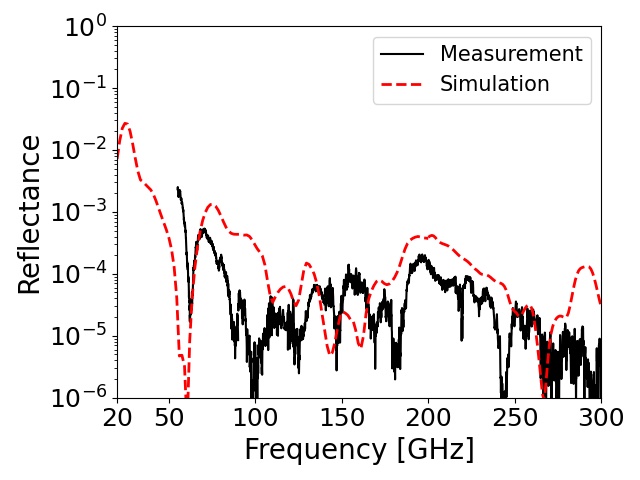}
            \subcaption{Carbon fiber (K223HE, 0.15\%)}
    \end{minipage}
    \begin{minipage}{0.45\textwidth}
            \includegraphics[width=0.9\textwidth]
            {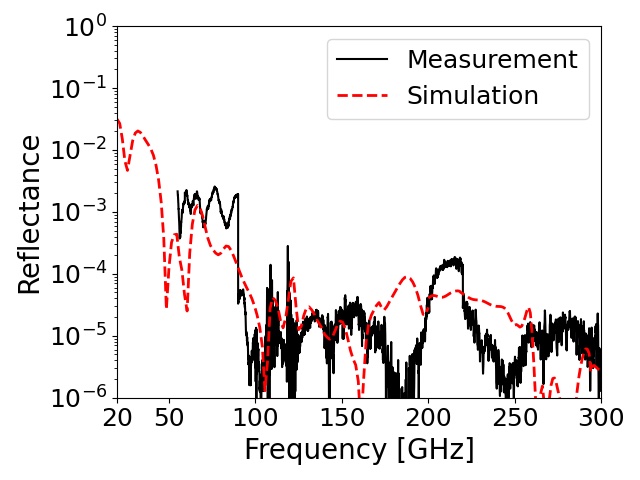}
            \subcaption{Carbon nanotube (SG101, 0.2\%)}
    \end{minipage}
    \caption{Reflectance of the RAM-3pm made with selected materials: mixture of STYCAST-2850FT and each absorptive material. The simulation results are broadly similar to the measurement.}
    \label{Reflectance_somesamples_simulation}
\end{figure*}

We fabricated four RAM-3pm with the selected materials:
the stainless steel powder (RD19-3532), two carbon fibers (K223HM and K223HE), and the carbon nanotube (SG101).
We measured their reflectance using the VNA in the frequency range of 50--300~GHz at room temperature.
The incident signal was polarized perpendicular to the incident plane (S-polarized).
Its incident angle to the RAM sample was $45^{\circ}$,
and the direction of the reflection signal was $45^{\circ}$. 

Figure~\ref{Reflectance_somesamples_simulation} shows the measured reflectance for each RAM-3pm.
In the range of 70--300~GHz, all of the samples achieved a low reflectance below $0.3\%$.
However, we observed the difference among the samples in the 50--70~GHz range.
The two samples made with K223HE or SG101 maintained a low reflectance, while the others did not.
We confirmed the statement described in the section~II and Fig.~\ref{k_comparison}; the reflectance relies on $\kappa$ at the frequency range below 70~GHz.

We also performed a simulation based on the configuration of the VNA measurements.
We used the measured parameters from the previous section, i.e., $n$ and $\kappa$ at the room temperature.
The simulation results are overlaid on the measured reflectance, as shown in Fig.~\ref{Reflectance_somesamples_simulation}.
They were broadly similar to the VNA measurements. 
Even below 50~GHz, the two samples mixed with K223HE or K223HM still maintained a low reflectance ($\lesssim 3\%$) according to the simulation.

\section{Discussion}

\begin{figure}[htpb]
    \centering
    \includegraphics[width=8cm]
    {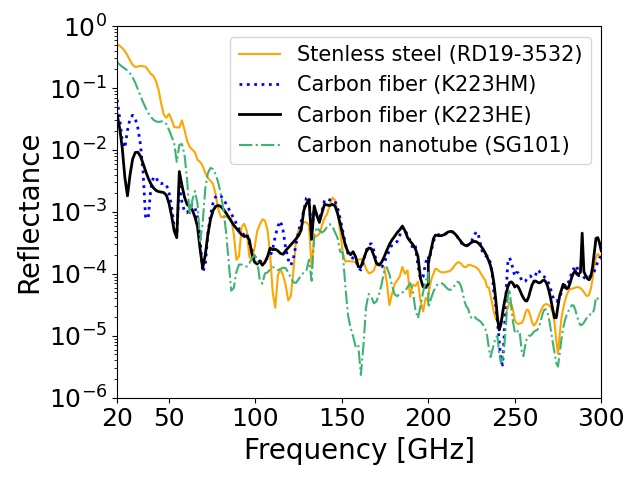}
    \caption{Results of the reflectance simulation for the 77~K condition. This simulation includes effects of the scattering at the surface and reflection at the backside of the RAM-3pm.}
    \label{copmarison_77K_ave}
\end{figure}

To understand the performance in a cryogenic application, we also simulated the reflectance at 77~K by using the measured parameters in the liquid nitrogen bath.
In this simulation study, we used the unpolarized signal as the real application.
We also included the scattered signal in any direction as well as the specular reflectance in the $45^{\circ}$ direction.
Figure~\ref{copmarison_77K_ave} shows the simulation results for each candidate materials in frequency range of 20--300~GHz.

A difference in the achieved reflectance among the four materials was noticeable below 70~GHz.
This tendency was implied by the room temperature study in the previous section.
The RAM-3pm made with the carbon fiber K223HE achieved the best performance.
This material also satisfied our requirement except at the lower edge of the frequency range~($\sim$20~GHz).
For the RAM-3pm made with carbon nanotube SG101, 
we observed a significant increase of the reflectance owing to the degradation of $\kappa$ at 77~K.

\begin{figure}[htpb]
    \centering
    \includegraphics[width=6cm]
    {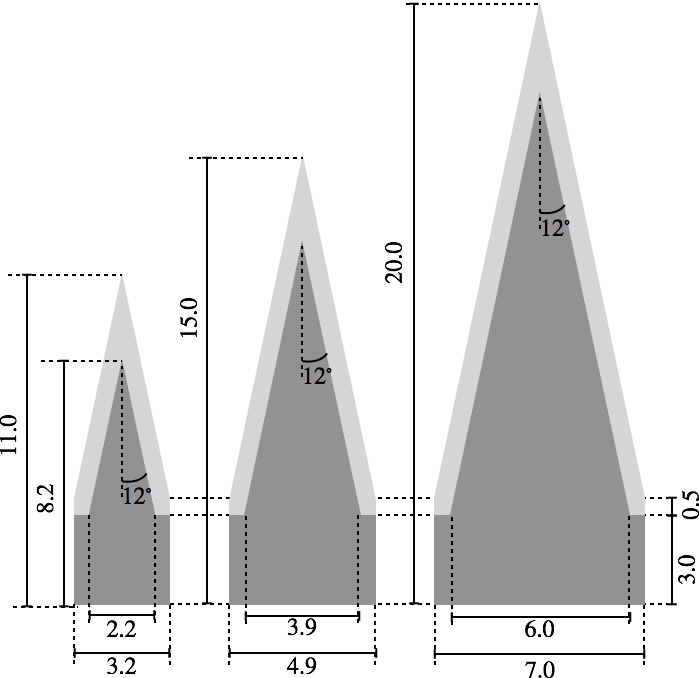}
    \\
    \includegraphics[width=8cm]
    {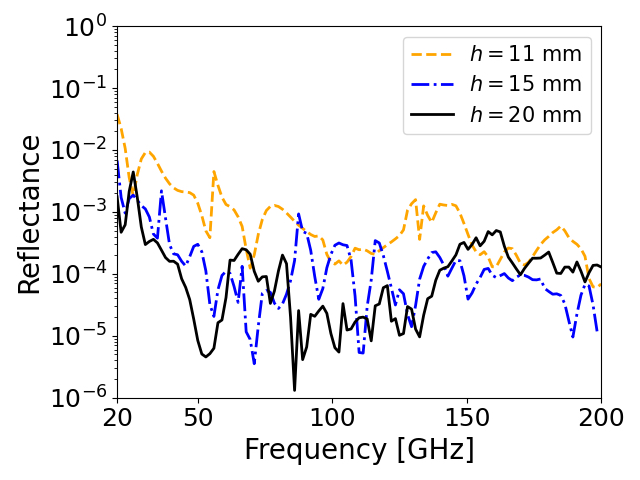}
    \caption{Results of the reflectance simulation for the RAM-3pm made with a mixture of STYCAST-2850FT and carbon fiber K223HE.
    We vary the geometry of the pyramidal structure.
    We use the optical parameters measured at 77~K.
    }
    \label{pyramid_change_simulation}
\end{figure}

To improve the reflectance around 20~GHz, we optimized the surface structure of the RAM-3pm. 
Figure~\ref{pyramid_change_simulation} shows simulation results with changing the height of the pyramidal structure.
We obtained a reflectance below 1\% when we double its height.
We also simulated the case of changing the thickness behind the pyramids. 
However, this did not result in any improvement.

\section{Conclusions}

We previously established the method to fabricate a radio absorptive material by using a 3D-printed mold, which is named as RAM-3pm.
The wide range of choices for materials is the advantage of this method compared with the other methods.
In this paper, we survey the best material for the RAM-3pm.
We set the requirement for the reflectance towards a CMB application:
$\lesssim 1\%$ in the wide frequency range of 20--300~GHz and in a cryogenic condition.
We select four materials based on the measured optical parameters ($n$ and $\kappa$) from 17 materials.
We made four RAM-3pm for each absorptive material, and we measured the reflectance for each of them at the room-temperature condition.
We confirm that the simulation results are broadly similar to the measurements.

We also simulate the reflectance at the low-temperature condition using the optical parameters obtained in the liquid nitrogen bath.
We conclude that the RAM-3pm made with the mixture of STYCAST-2850FT and carbon fiber K223HE achieves the best performance.
This satisfies our requirement except for around the 20~GHz range.

We discuss how to improve the reflectance by changing the surface structure of the RAM-3pm.
We estimate that it is possible to achieve a reflectance below 1\% in the frequency range of 20--300~GHz~\
when we double the height of the pyramidal structure.
These estimations will be confirmed by using an improved system for the reflectance measurement at 20--50~GHz.
Understanding the performance at even lower temperatures ($\sim\mathrm{K}$) is also a future project.

\section*{Acknowledgements}

We thank 
Toyo Aluminium K.K., 
Ito Graphite Co. Ltd., 
Mitsubishi Chemical Corp., 
ALMEDIO INC., 
Zeon Nano Technology Co. Ltd., 
Henkel AG $\&$ Co. KGaA, 
E$\&$C Engineering K.K., 
and CEMEDINE CO. LTD. 
for providing us with the absorptive materials which were used in this paper.
This work is supported by JSPS KAKENHI under grant numbers JP17H06134 and JP21H00071, 
and also supported by World Premier International Research Center Initiative (WPI), MEXT, Japan.
TO acknowledges the JSPS core-to-core program JPJSCCA20200003 for his travel supports.
SA also acknowledges JP18J01039.
We thank Tomotake Matsumura and Edanz (https://jp.edanz.com/ac) for editing a draft of this manuscript.

\section*{Data Availability}

The data that support the findings of this study are available from the corresponding author upon reasonable request.

\nocite{*}
\bibliography{main}

\providecommand{\noopsort}[1]{}\providecommand{\singleletter}[1]{#1}%
\begin{thebibliography}{15}%
\makeatletter
\providecommand \@ifxundefined [1]{%
 \@ifx{#1\undefined}
}%
\providecommand \@ifnum [1]{%
 \ifnum #1\expandafter \@firstoftwo
 \else \expandafter \@secondoftwo
 \fi
}%
\providecommand \@ifx [1]{%
 \ifx #1\expandafter \@firstoftwo
 \else \expandafter \@secondoftwo
 \fi
}%
\providecommand \natexlab [1]{#1}%
\providecommand \enquote  [1]{``#1''}%
\providecommand \bibnamefont  [1]{#1}%
\providecommand \bibfnamefont [1]{#1}%
\providecommand \citenamefont [1]{#1}%
\providecommand \href@noop [0]{\@secondoftwo}%
\providecommand \href [0]{\begingroup \@sanitize@url \@href}%
\providecommand \@href[1]{\@@startlink{#1}\@@href}%
\providecommand \@@href[1]{\endgroup#1\@@endlink}%
\providecommand \@sanitize@url [0]{\catcode `\\12\catcode `\$12\catcode
  `\&12\catcode `\#12\catcode `\^12\catcode `\_12\catcode `\%12\relax}%
\providecommand \@@startlink[1]{}%
\providecommand \@@endlink[0]{}%
\providecommand \url  [0]{\begingroup\@sanitize@url \@url }%
\providecommand \@url [1]{\endgroup\@href {#1}{\urlprefix }}%
\providecommand \urlprefix  [0]{URL }%
\providecommand \Eprint [0]{\href }%
\providecommand \doibase [0]{http://dx.doi.org/}%
\providecommand \selectlanguage [0]{\@gobble}%
\providecommand \bibinfo  [0]{\@secondoftwo}%
\providecommand \bibfield  [0]{\@secondoftwo}%
\providecommand \translation [1]{[#1]}%
\providecommand \BibitemOpen [0]{}%
\providecommand \bibitemStop [0]{}%
\providecommand \bibitemNoStop [0]{.\EOS\space}%
\providecommand \EOS [0]{\spacefactor3000\relax}%
\providecommand \BibitemShut  [1]{\csname bibitem#1\endcsname}%
\let\auto@bib@innerbib\@empty
\bibitem [{\citenamefont {Ade}\ \emph {et~al.}(2019)\citenamefont {Ade} \emph
  {et~al.}}]{bib:so}%
  \BibitemOpen
  \bibfield  {author} {\bibinfo {author} {\bibfnamefont {P.}~\bibnamefont
  {Ade}} \emph {et~al.},\ }\bibfield  {title} {\enquote {\bibinfo {title} {The
  simons observatory: science goals and forecasts},}\ }\href {\doibase
  10.1088/1475-7516/2019/02/056} {\bibfield  {journal} {\bibinfo  {journal}
  {Journal of Cosmology and Astroparticle Physics}\ }\textbf {\bibinfo {volume}
  {2019}},\ \bibinfo {pages} {056--056} (\bibinfo {year} {2019})}\BibitemShut
  {NoStop}%
\bibitem [{\citenamefont {Adachi}\ \emph {et~al.}(2020)\citenamefont {Adachi},
  \citenamefont {Hattori}, \citenamefont {Kanno}, \citenamefont {Kiuchi},
  \citenamefont {Okada},\ and\ \citenamefont {Tajima}}]{bib:Adachi_paper}%
  \BibitemOpen
  \bibfield  {author} {\bibinfo {author} {\bibfnamefont {S.}~\bibnamefont
  {Adachi}}, \bibinfo {author} {\bibfnamefont {M.}~\bibnamefont {Hattori}},
  \bibinfo {author} {\bibfnamefont {F.}~\bibnamefont {Kanno}}, \bibinfo
  {author} {\bibfnamefont {K.}~\bibnamefont {Kiuchi}}, \bibinfo {author}
  {\bibfnamefont {T.}~\bibnamefont {Okada}}, \ and\ \bibinfo {author}
  {\bibfnamefont {O.}~\bibnamefont {Tajima}},\ }\bibfield  {title} {\enquote
  {\bibinfo {title} {Production method of millimeter-wave absorber with
  3d-printed mold},}\ }\href {\doibase 10.1063/1.5132871} {\bibfield  {journal}
  {\bibinfo  {journal} {Review of Scientific Instruments}\ }\textbf {\bibinfo
  {volume} {91}},\ \bibinfo {pages} {016103} (\bibinfo {year} {2020})},\
  \Eprint {http://arxiv.org/abs/https://doi.org/10.1063/1.5132871}
  {https://doi.org/10.1063/1.5132871} \BibitemShut {NoStop}%
\bibitem [{\citenamefont {Xu}\ \emph {et~al.}(2021)\citenamefont {Xu},
  \citenamefont {Chesmore}, \citenamefont {Adachi}, \citenamefont {Ali},
  \citenamefont {Bazarko}, \citenamefont {Coppi}, \citenamefont {Devlin},
  \citenamefont {Devlin}, \citenamefont {Dicker}, \citenamefont {Gallardo},
  \citenamefont {Golec}, \citenamefont {Gudmundsson}, \citenamefont
  {Harrington}, \citenamefont {Hattori}, \citenamefont {Kofman}, \citenamefont
  {Kiuchi}, \citenamefont {Kusaka}, \citenamefont {Limon}, \citenamefont
  {Matsuda}, \citenamefont {McMahon}, \citenamefont {Nati}, \citenamefont
  {Niemack}, \citenamefont {Suzuki}, \citenamefont {Teply}, \citenamefont
  {Thornton}, \citenamefont {Wollack}, \citenamefont {Zannoni},\ and\
  \citenamefont {Zhu}}]{bib:MMA}%
  \BibitemOpen
  \bibfield  {author} {\bibinfo {author} {\bibfnamefont {Z.}~\bibnamefont
  {Xu}}, \bibinfo {author} {\bibfnamefont {G.~E.}\ \bibnamefont {Chesmore}},
  \bibinfo {author} {\bibfnamefont {S.}~\bibnamefont {Adachi}}, \bibinfo
  {author} {\bibfnamefont {A.~M.}\ \bibnamefont {Ali}}, \bibinfo {author}
  {\bibfnamefont {A.}~\bibnamefont {Bazarko}}, \bibinfo {author} {\bibfnamefont
  {G.}~\bibnamefont {Coppi}}, \bibinfo {author} {\bibfnamefont
  {M.}~\bibnamefont {Devlin}}, \bibinfo {author} {\bibfnamefont
  {T.}~\bibnamefont {Devlin}}, \bibinfo {author} {\bibfnamefont {S.~R.}\
  \bibnamefont {Dicker}}, \bibinfo {author} {\bibfnamefont {P.~A.}\
  \bibnamefont {Gallardo}}, \bibinfo {author} {\bibfnamefont {J.~E.}\
  \bibnamefont {Golec}}, \bibinfo {author} {\bibfnamefont {J.~E.}\ \bibnamefont
  {Gudmundsson}}, \bibinfo {author} {\bibfnamefont {K.}~\bibnamefont
  {Harrington}}, \bibinfo {author} {\bibfnamefont {M.}~\bibnamefont {Hattori}},
  \bibinfo {author} {\bibfnamefont {A.}~\bibnamefont {Kofman}}, \bibinfo
  {author} {\bibfnamefont {K.}~\bibnamefont {Kiuchi}}, \bibinfo {author}
  {\bibfnamefont {A.}~\bibnamefont {Kusaka}}, \bibinfo {author} {\bibfnamefont
  {M.}~\bibnamefont {Limon}}, \bibinfo {author} {\bibfnamefont
  {F.}~\bibnamefont {Matsuda}}, \bibinfo {author} {\bibfnamefont
  {J.}~\bibnamefont {McMahon}}, \bibinfo {author} {\bibfnamefont
  {F.}~\bibnamefont {Nati}}, \bibinfo {author} {\bibfnamefont {M.~D.}\
  \bibnamefont {Niemack}}, \bibinfo {author} {\bibfnamefont {A.}~\bibnamefont
  {Suzuki}}, \bibinfo {author} {\bibfnamefont {G.~P.}\ \bibnamefont {Teply}},
  \bibinfo {author} {\bibfnamefont {R.~J.}\ \bibnamefont {Thornton}}, \bibinfo
  {author} {\bibfnamefont {E.~J.}\ \bibnamefont {Wollack}}, \bibinfo {author}
  {\bibfnamefont {M.}~\bibnamefont {Zannoni}}, \ and\ \bibinfo {author}
  {\bibfnamefont {N.}~\bibnamefont {Zhu}},\ }\bibfield  {title} {\enquote
  {\bibinfo {title} {The simons observatory: metamaterial microwave absorber
  and its cryogenic applications},}\ }\href {\doibase 10.1364/AO.411711}
  {\bibfield  {journal} {\bibinfo  {journal} {Appl. Opt.}\ }\textbf {\bibinfo
  {volume} {60}},\ \bibinfo {pages} {864--874} (\bibinfo {year}
  {2021})}\BibitemShut {NoStop}%
\bibitem [{\citenamefont {Petroff}\ \emph {et~al.}(2019)\citenamefont
  {Petroff}, \citenamefont {Appel}, \citenamefont {Rostem}, \citenamefont
  {Bennet~t}, \citenamefont {Eimer}, \citenamefont {Marriage}, \citenamefont
  {Ramirez},\ and\ \citenamefont {Wollack}}]{bib:EdWollack}%
  \BibitemOpen
  \bibfield  {author} {\bibinfo {author} {\bibfnamefont {M.}~\bibnamefont
  {Petroff}}, \bibinfo {author} {\bibfnamefont {J.}~\bibnamefont {Appel}},
  \bibinfo {author} {\bibfnamefont {K.}~\bibnamefont {Rostem}}, \bibinfo
  {author} {\bibfnamefont {C.~L.}\ \bibnamefont {Bennet~t}}, \bibinfo {author}
  {\bibfnamefont {J.}~\bibnamefont {Eimer}}, \bibinfo {author} {\bibfnamefont
  {T.}~\bibnamefont {Marriage}}, \bibinfo {author} {\bibfnamefont
  {J.}~\bibnamefont {Ramirez}}, \ and\ \bibinfo {author} {\bibfnamefont
  {E.~J.}\ \bibnamefont {Wollack}},\ }\bibfield  {title} {\enquote {\bibinfo
  {title} {A 3d-printed broadband millimeter wave absorber},}\ }\href {\doibase
  10.1063/1.5050781} {\bibfield  {journal} {\bibinfo  {journal} {Review of
  Scientific Instruments}\ }\textbf {\bibinfo {volume} {90}},\ \bibinfo {pages}
  {024701} (\bibinfo {year} {2019})},\ \Eprint
  {http://arxiv.org/abs/https://doi.org/10.1063/1.5050781}
  {https://doi.org/10.1063/1.5050781} \BibitemShut {NoStop}%
\bibitem [{\citenamefont {{ANSYS, Inc.}}()}]{bib:ANSYS}%
  \BibitemOpen
  \bibfield  {author} {\bibinfo {author} {\bibnamefont {{ANSYS, Inc.}}},\
  }\href {https://www.ansys.com/} {}\bibinfo {note} {Southpointe, 2600 Ansys
  Drive, Canonsburg, PA 15317, USA},\ \Eprint
  {http://arxiv.org/abs/https://www.ansys.com/} {https://www.ansys.com/}
  \BibitemShut {NoStop}%
\bibitem [{\citenamefont {Jackson}(1999)}]{bib:jackson}%
  \BibitemOpen
  \bibfield  {author} {\bibinfo {author} {\bibfnamefont {J.~D.}\ \bibnamefont
  {Jackson}},\ }\href {http://cdsweb.cern.ch/record/490457} {\emph {\bibinfo
  {title} {Classical electrodynamics}}},\ \bibinfo {edition} {3rd}\ ed.\
  (\bibinfo  {publisher} {Wiley},\ \bibinfo {address} {New York, {NY}},\
  \bibinfo {year} {1999})\BibitemShut {NoStop}%
\bibitem [{\citenamefont {{Henkel Japan Ltd.}}()}]{bib:Henkel}%
  \BibitemOpen
  \bibfield  {author} {\bibinfo {author} {\bibnamefont {{Henkel Japan Ltd.}}},\
  }\href {https://www.henkel.com} {}\bibinfo {note} {14F Sphere Tower Tennozu,
  2-2-8 Higashi Shinagawa, Shinagawa-ku, Tokyo, 140-0002, Japan},\ \Eprint
  {http://arxiv.org/abs/https://www.henkel.com} {https://www.henkel.com}
  \BibitemShut {NoStop}%
\bibitem [{\citenamefont {{Toyo Aluminium K.K.}}()}]{bib:ToyoAluminium}%
  \BibitemOpen
  \bibfield  {author} {\bibinfo {author} {\bibnamefont {{Toyo Aluminium
  K.K.}}},\ }\href {https://www.toyal.co.jp/eng/} {}\bibinfo {note} {6-8,
  Kyutaromachi 3-chome, Chuo-ku, Osaka 541-0056 JAPAN},\ \Eprint
  {http://arxiv.org/abs/https://www.toyal.co.jp/eng/}
  {https://www.toyal.co.jp/eng/} \BibitemShut {NoStop}%
\bibitem [{\citenamefont {{Ito Graphite Co., Ltd.}}()}]{bib:ItoGraphite}%
  \BibitemOpen
  \bibfield  {author} {\bibinfo {author} {\bibnamefont {{Ito Graphite Co.,
  Ltd.}}},\ }\href {http://www.graphite.co.jp/index.html} {}\bibinfo {note}
  {970, Tado-chou Tado, Kuwana-shi, Mie 511-0106 JAPAN},\ \Eprint
  {http://arxiv.org/abs/http://www.graphite.co.jp/index.html}
  {http://www.graphite.co.jp/index.html} \BibitemShut {NoStop}%
\bibitem [{\citenamefont {{Mitsubishi Chemical
  Corporation}}()}]{bib:MCCarbonBlack}%
  \BibitemOpen
  \bibfield  {author} {\bibinfo {author} {\bibnamefont {{Mitsubishi Chemical
  Corporation}}},\ }\href {http://www.carbonblack.jp/en/index.html} {}\bibinfo
  {note} {1-1 Marunouchi 1-chome, Chiyoda-ku, Tokyo 100-8251, Japan},\ \Eprint
  {http://arxiv.org/abs/http://www.carbonblack.jp/en/index.html}
  {http://www.carbonblack.jp/en/index.html} \BibitemShut {NoStop}%
\bibitem [{\citenamefont {{ALMEDIO INC.}}()}]{bib:ALMEDIO}%
  \BibitemOpen
  \bibfield  {author} {\bibinfo {author} {\bibnamefont {{ALMEDIO INC.}}},\
  }\href {https://www.almedio.co.jp/en/} {}\bibinfo {note} {KS Kunitachi Bld.
  7F,1-4-12, Higashi,Kunitachi-shi, Tokyo,186-0002, Japan},\ \Eprint
  {http://arxiv.org/abs/https://www.almedio.co.jp/en/}
  {https://www.almedio.co.jp/en/} \BibitemShut {NoStop}%
\bibitem [{\citenamefont {{Mitsubishi Chemical Corporation
  }}()}]{bib:MCCarbonFiber}%
  \BibitemOpen
  \bibfield  {author} {\bibinfo {author} {\bibnamefont {{Mitsubishi Chemical
  Corporation }}},\ }\href
  {https://www.m-chemical.co.jp/en/products/departments/mcc/cfcm/product/1201229\_7502.html}
  {}\bibinfo {note} {1-1 Marunouchi 1-chome, Chiyoda-ku, Tokyo 100-8251,
  Japan},\ \Eprint
  {http://arxiv.org/abs/https://www.m-chemical.co.jp/en/products/departments/mcc/cfcm/product/1201229\_7502.html}
  {https://www.m-chemical.co.jp/en/products/departments/mcc/cfcm/product/1201229\_7502.html}
  \BibitemShut {NoStop}%
\bibitem [{\citenamefont {{Zeon Nano Technology Co.,
  Ltd.}}()}]{bib:ZeonNanoTechnology}%
  \BibitemOpen
  \bibfield  {author} {\bibinfo {author} {\bibnamefont {{Zeon Nano Technology
  Co., Ltd.}}},\ }\href {http://www.zeonnanotech.jp/en/index.html} {}\bibinfo
  {note} {Shin Marunouchi Center Building 14th Floor, 1-6-2 Marunouchi,
  Chiyoda-ku, Tokyo Japan},\ \Eprint
  {http://arxiv.org/abs/http://www.zeonnanotech.jp/en/index.html}
  {http://www.zeonnanotech.jp/en/index.html} \BibitemShut {NoStop}%
\bibitem [{\citenamefont {Ohta}, \citenamefont {Hattori},\ and\ \citenamefont
  {Matsuo}(2006)}]{bib:HattoriFTS}%
  \BibitemOpen
  \bibfield  {author} {\bibinfo {author} {\bibfnamefont {I.~S.}\ \bibnamefont
  {Ohta}}, \bibinfo {author} {\bibfnamefont {M.}~\bibnamefont {Hattori}}, \
  and\ \bibinfo {author} {\bibfnamefont {H.}~\bibnamefont {Matsuo}},\
  }\bibfield  {title} {\enquote {\bibinfo {title} {{Development of a
  multi-Fourier-transform interferometer: fundamentals}},}\ }\href {\doibase
  10.1364/AO.45.002576} {\bibfield  {journal} {\bibinfo  {journal} {Appl.
  Opt.}\ }\textbf {\bibinfo {volume} {45}},\ \bibinfo {pages} {2576--2585}
  (\bibinfo {year} {2006})}\BibitemShut {NoStop}%
\bibitem [{\citenamefont {{Ohta}}\ \emph {et~al.}(2007)\citenamefont {{Ohta}},
  \citenamefont {{M. Hattori}}, \citenamefont {{Y. Chinone}}, \citenamefont
  {{Y. Luo}}, \citenamefont {{Y. Hamaji}}, \citenamefont {{J. Takahashi}},
  \citenamefont {{H. Matsuo}},\ and\ \citenamefont {{N.
  Kuno}}}]{bib:HattoriBolometer}%
  \BibitemOpen
  \bibfield  {author} {\bibinfo {author} {\bibfnamefont {I.~S.}\ \bibnamefont
  {{Ohta}}}, \bibinfo {author} {\bibnamefont {{M. Hattori}}}, \bibinfo {author}
  {\bibnamefont {{Y. Chinone}}}, \bibinfo {author} {\bibnamefont {{Y. Luo}}},
  \bibinfo {author} {\bibnamefont {{Y. Hamaji}}}, \bibinfo {author}
  {\bibnamefont {{J. Takahashi}}}, \bibinfo {author} {\bibnamefont {{H.
  Matsuo}}}, \ and\ \bibinfo {author} {\bibnamefont {{N. Kuno}}},\ }\bibfield
  {title} {\enquote {\bibinfo {title} {{Astronomical testing observation in
  multi-Fourier transform interferometer: Aperture synthesis technique and
  CMB}},}\ }in\ \href {\doibase 10.1109/ICIMW.2007.4516522} {\emph {\bibinfo
  {booktitle} {2007 Joint 32nd International Conference on Infrared and
  Millimeter Waves and the 15th International Conference on Terahertz
  Electronics}}}\ (\bibinfo {year} {2007})\ pp.\ \bibinfo {pages}
  {335--336}\BibitemShut {NoStop}%
\end{thebibliography}%

\end{document}